\documentclass[12pt]{article}           
\usepackage{amstex}
\renewcommand{\tt}{\scriptscriptstyle}
\def\VR#1#2{\vrule height #1mm depth #2mm width 0pt}
\def\TVR#1#2{@{~~\VR{#1}{#2}}}

\def\plb#1 #2 {Phys. Lett. {\bf #1B} #2 }
\def\phr#1 #2 {Phys. Rep. {\bf  #1} #2 }        
\def\npb#1 #2 {Nucl. Phys. {\bf B#1} #2 }
\def\aph#1 #2 {Ann. Phys. {\bf #1} #2 }         
\def\jmp#1 #2 {J. Math. Phys. {\bf #1} #2 }
\def\jgp#1 #2 {J. Geom. Phys. {\bf #1} #2 }
\def\prd#1 #2 {Phys. Rev. {\bf D#1} #2 }
\def\prl#1 #2 {Phys. Rev. Lett. {\bf #1} #2 }
\def\rmp#1 #2 {Rev. Mod. Phys.  {\bf #1} #2 }
\def\zpc#1 {Z. Phys. {\bf #1C} }
\def\cmp#1 #2 {Commun. Math. Phys. {\bf #1} #2 }
\def\cqg#1 #2 {Class.Quant.Grav. {\bf #1} #2 }
\def\mpl#1 {Mod. Phys. Lett. {\bf A#1} }
\def\cpc#1 {Computer Phys. Commun. {\bf #1} }   
\def\ijmp#1 {Int. J. Mod. Phys. {\bf A#1} }
\def\ijmpC#1 {Int. J. Mod. Phys. {\bf C#1} }

\begin{document}

\begin{flushright}
hep-th/9612067\\
TUW--96--30
\end{flushright}

\vspace{20mm}

\begin{center}
{\LARGE\bf On the Calabi--Yau Phase of $(0,2)$ Models}
\end{center}

\vspace{15mm}

\begin{center} \vskip 12mm
  
{
\large Mahmoud Nikbakht--Tehrani
\footnote[2]{e-mail: \texttt{nikbakht@@tph16.tuwien.ac.at}}}

\vskip 5mm
       Institut f\"{u}r Theoretische Physik, Technische Universit\"{a}t Wien\\
       Wiedner Hauptstra\ss{}e 8--10, A-1040 Wien, AUSTRIA
\end{center}

\vspace{20mm}

\begin{center}
\begin{minipage}{12cm}
\begin{center}
{\bf Abstract}
\end{center}
We study the Calabi-Yau phase of a certain class of $(0,2)$ 
models. These are conjectured to be equivalent to exact $(0,2)$ 
superconformal field theories which have been constructed recently. Using
the methods of toric geometry we discuss in a few examples 
the problem of resolving the singularities of
such models and calculate the Euler characteristic of the corresponding 
gauge bundles. 
\end{minipage}
\end{center}

\vspace{15mm}

December 1996
\thispagestyle{empty}


\newpage

\section{Introduction}

 The classical solutions to the purturbative string theory with unbroken
$N=1$ spacetime supersymmetry provide us with the only known string 
vacua in four dimensions. As is well-known, the $(0,2)$
superconformal invariance on the string worldsheet together with 
an integrality condition on the charges of $U(1)$ current in 
the superconformal algebra are equivalent to $N=1$ spacetime 
supersymmetry \cite{ba88,ge88}. Therefore, the $(0,2)$ superconformal field
theories (particularly the $(0,2)$ Calabi-Yau
$\sigma$ models) seem to be the natural context for the (geometric) 
string compactification. Yet another source of interest in $(0,2)$ models is
due to the fact that the phenomenological prospects of such models are much more
promising than those of, e.g. $(2,2)$ theories, because 
they lead to the more realistic gauge groups \cite{hu85,wi86} 
like $SO(10)$ and $SU(5)$.\\

Nevertheless, these models have received less
attention in the early days of string theory. This was in part because of
the assertion made in \cite{di86} that the generic $(0,2)$ Calabi-Yau
$\sigma$ models suffer from destabilization by the worldsheet 
instantons. (However, the recent work of \cite{si95} shows that 
these models are not destabilized by such nonperturbative $\sigma$
model effects.)
In spite of the early work of \cite{di871,di881,gr90} the technical difficulty 
in constructing $(0,2)$ models remained  the other
main obstacle in the study of these models. \\

The Witten's gauged linear sigma model approach \cite{wi93} has dramatically 
changed the state of affairs.
It provided a powerful tool in constructing $(2,2)$ and $(0,2)$ 
models and in analyzing their `phase structures'. Using this framework
the authors of \cite{di94} have constructed and analyzed plenty 
of $(0,2)$ models in their `Landau-Ginzburg phase'. The subsequent works of 
\cite{bl95,bl96} proposed an identification between an exact $(0,2)$ SCFT and
a certain model of \cite{di94}. This exact $(0,2)$ SCFT was a Gepner 
type model which was constructed  using the simple current methods. Inspired
by this proposal the authors of \cite{bl961,nik961} have tried to extend this 
identification to a larger class of $(0,2)$ models. The starting point 
was a general solution of anomaly cancellation condition 
yielding a large set of consistent $\sigma$ model data.
In these works the attention has been paid primarily to the 
connection of the exact SCFTs and the Landau-Ginzburg models.\\

The above mentioned geometric data defining a $(0,2)$ supersymmetric 
$U(1)$ gauge theory in general result in a singular 
Calabi-Yau variety and some stable vector bundle ($\simeq$ locally
free sheaf), the so-called gauge bundle (or `gauge sheaf'), on it. 
The `Calabi-Yau 
phases' correspond to the possible crepant desingularizations of this singular 
model. As discussed in \cite{di96}, the $(0,2)$ singularities have two different
origins. One set of them comes from the singularities of the base Calabi-Yau
variety and the other one  is associated to the singularities of the
gauge sheaf. (These are points of the base varieties, where the gauge sheaf
fails to be locally free.)\\

In this paper we are going to study the Calabi-Yau 
phase of some examples from the class of models 
constructed in \cite{bl961,nik962}. 
We take the same 
attitude towards this problem as the authors of \cite{di96}. 
After resolving the
singularities of the  base variety  we calculate the
Euler characteristic of the `pulled back gauge bundles'. The methods which
are used here are those of 
toric geometry, specially the intersection theory and
the Riemann-Roch theorem for coherent sheaves on the toric varieties.\\

In section 2 we briefly review some basic aspects of the gauged linear
sigma model approach and give the solution of the anomaly cancellation
condition found in \cite{bl961,nik961}. 
As mentioned above, this general solution
provides a large set of consistent data for $(0,2)$ Calabi-Yau $\sigma$ models.
Using these data we construct the models of 
interest to us. The next section will be of 
a technical nature. Here we discuss the relevant 
mathematical tools from toric geometry. In section 4 we discuss 
our examples. We conclude with some comments about open 
problems and directions for future work.


\section{The gauged linear sigma models}

In this section  we explain the basic 
ideas of the gauged linear sigma model approach  without going 
into details. In doing so we pinpoint those aspects which are
crucial for our considerations in the next sections. For more
details we refer to \cite{wi93,di94,di95}. \\

The starting point is a $(0,2)$ supersymmetric $U(1)$ gauge 
theory that represents a nonconformal member of the universality class 
of a  $(0,2)$ superconformal field theory. The action $S$ which describes 
this model is
\begin{eqnarray}
S=S_{g}+S_{k}+S_{\tt W}+S_{\tt F.I.}\;\; , 
\end{eqnarray}
where $S_{g}$, $ S_{k}$ are the kinetic parts of the gauge
and matter sectors, $S_{\tt W}$ is the superpotential and 
$S_{\tt F.I.}$ is the $U(1)$  Fayet-Iliopoulos D-term.\\

Let $\Phi_{i} (i=1,\ldots ,n+1) $, $P$ be chiral 
scalar superfields with the $U(1)$ charges $w_{i}, -m$ and $\Lambda^{a} 
(a=1,\ldots ,\ell +1) $, $\Gamma$ be  chiral Fermi superfields with the
$U(1)$ charges $q_{a}, -d$.
The simplest $(0,2)$ superpotential that one can write down has the 
follwoing form  
\begin{eqnarray}
S_{\tt W}=\int d^{2}z\, d\theta \;\;\left(\Gamma W(\Phi_{i})+
                                P\Lambda^{a}F_{a}(\Phi_{i})\right),
\end{eqnarray}
where $W$ and $F_{a}$ are homogeneous polynomials in $\Phi_{i}$ of
degree $d$ and $m-q_{a}$, respectively.
Integrating out the D auxiliary field in the gauge multiplet and the 
auxiliary fields in the chiral Fermi superfields, we get the bosonic
potential
\begin{eqnarray}
U=|W(\phi_{i})|^{2}+|p|^{2}\;\sum_{a}|F_{a}(\phi_{i})|^{2}+
\frac{e^{2}}{2}\left(\sum_{i}w_{i}|\phi_{i}|^{2}-m|p|^{2}-r\right)^{2}
\end{eqnarray}
where the parameter $r$ is the coefficient in the   
Fayet-Iliopoulos D-term and $\phi_{i}, p$ denote the lowest terms of the
superfields $\Phi_{i}$ and $P$.\\

Now varying the parameter $r$ this model exhibits different `phases'. By 
minimizing the classical bosonic potential $U$ for
large positive $r$ we obtain 
\begin{eqnarray}
\sum_{i}w_{i}|\phi_{i}|^{2}=r,\;\;\; W(\phi_{i})=0,\;\;\; p=0\; .
\end{eqnarray}
Taking the quotient by the action of the $U(1)$ gauge group  
these equations describe a Calabi-Yau variety $X$ as the zero locus 
of the homogeneous polynomial
$W(\phi_{i})$ in the weighted projective space ${\Bbb P}
(w_{1},\ldots ,w_{n+1})$ 
with the K\"{a}hler class proportional to $r$.
The right-moving fermions $\psi_{i}$ (the superpartners of $\phi_{i}$) 
couple to the tangent bundle of $X$ which is given by the cohomology
of the monad
\begin{eqnarray} 
0\rightarrow {\cal O}\rightarrow \bigoplus_{i=1}^{n+1}
{\cal O}(w_{i})\rightarrow {\cal O}(d)\rightarrow 0\label{1a}, 
\end{eqnarray}
and in the same way the left-moving fermions $\lambda_{a}$ 
(the lowest components of the superfields $\Lambda_{a}$) couple to the
vector bundle $V$ defined by the cohomology of the following monad
\begin{eqnarray} 
0\rightarrow {\cal O}\rightarrow \bigoplus_{a=1}^{\ell +1}
{\cal O}(q_{a})\rightarrow {\cal O}(m)\rightarrow 0. 
\end{eqnarray}
So we find that our gauged linear
sigma model for positive $r$ reduces in the infrared limit to a $(0,2)$ 
Calabi-Yau $\sigma$ model with the target space $X$, a hypersurface in
the weighted projective space ${\Bbb P}(w_{1},\ldots ,w_{n+1})$, and 
a rank $\ell$ gauge bundle $V$ on it which is defined by 
the exact sequence
\begin{eqnarray} 
0\rightarrow V \rightarrow \bigoplus_{a=1}^{\ell+1}
{\cal O}(q_{a})\stackrel{F_{a}}{\longrightarrow} {\cal O}(m)\rightarrow 0. 
\end{eqnarray}     
But this is not the whole story! Apart from some `regularity' conditions on
$F_{a}$ and $q_{a}$ \cite{di94}, these geometric data still have to 
satisfy an important condition that comes from the cancellation of 
the $U(1)$ gauge anomaly. 
Imposing the condition $c_{2}(V)=c_{2}(X)$ guarantees this cancellation. 
This leads, in turn, to the following quadratic Diophantine 
equation:
\begin{equation}
      m^{2}-\sum_{a=1}^{\ell+1} q_{a}^{2}=d^{2}
                                       -\sum_{i=1}^{n+1}w_{i}^{2}\;\;\label{1}.
\end{equation} 
Note also that, with the above choice of $U(1)$ charges, the first 
Chern class of $V$ vanishes
\begin{equation}
             \sum_{a=1}^{\ell+1} q_{a}- m=0 \;\;.
\end{equation}
which guarantees the existence of 
spinors on it. For large negative values of $r$ the vanishing of the 
bosonic potential yields   
\begin{equation}
 \phi_{i}=0 \;\;\; (\;\mbox{for all}\;\; i\;\;), 
 \;\; |p|^{2}=\frac{-r}{m} \;\;.
\end{equation}
In this case $p$ and
its superpartner become massive and drop out of the low energy theory. 
The gauge group $U(1)$ breaks down to the subgroup
${\Bbb Z}_{m}$ because the charge of $p$ is $m$.
We are therefore left with a $(0,2)$ Landau-Ginzburg 
orbifold in the infrared limit.
Absorbing the vacuum expectation value of $p$ by a trivial rescaling of the
fields we get the superpotential
\begin{eqnarray}
S_{\tt W}=\int d^{2}z\, d\theta \;\;\left(\Gamma W(\Phi_{i})+
                                \Lambda^{a}F_{a}(\Phi_{i})\right).
\end{eqnarray} 
Summarizing the above discussion we have found that 
the Calabi-Yau $\sigma$ models and the Landau-Ginzburg models can be
interpreted as two different `phases' 
of the same underlying theory.\\

It should be noted that the Calabi-Yau varieties of interest, 
which have been defined as hypersurfaces in weighted projective spaces 
are in general singular, whereas their corresponding 
physical theories are well-behaved. This is an indication of the fact
that the strings probe the smooth geometry of the target Calabi-Yau
spaces. Therefore, our phase picture of the moduli space of the theory is
not complete. To remedy this we have to desingularize our model and consider
the moduli space of this new model within which the moduli space of the 
original model will be embedded.\\

In the framework of gauged linear sigma
models the process of desingularization amounts above all to embedding 
the original theory in a new one with gauge group 
$U(1)\times\ldots \times U(1)$ 
($N$ copies) and $N-1$  new chiral scalar superfields $\Upsilon_{1},
\ldots , \Upsilon_{N-1}$ 
and then determine the charges of the fields 
with respect to the full gauge group. 
This new model 
then has $N$ K\"{a}hler moduli parameters 
$r_{1},\ldots, r_{\tt N}$, one for each Fayet-Iliopoulos 
D-term of the $U(1)$ factors of the new gauge group. Now by varying 
these parameters and finding the minima of the bosonic potential one
can recover as before the phase structure of the moduli space.\\
 
As is well-known \cite{as941,as942}, there is an equivalent formulation 
of the whole story in terms of toric geometry. It provides
us with some efficient computational tools for 
analyzing the phase structure of the
moduli space. The basic idea here is that the relevant information describing
a theory is encoded in the combinatorial data of some reflexive 
polytope $\Delta$ in a lattice $\bf N$ and the phase 
structure of the theory is then determined by the possible 
triangulations of this polytope. By the Calabi-Yau phase we now 
mean a phase which corresponds to a maximal triangulation of $\Delta$ .
We still have to deal with 
the other set of data in a $(0,2)$ model, namely
the gauge bundle data.  
We postpone this issue and the technicalities of resolving the 
base variety with the methods of toric geometry  to the next section.\\

Now we give a general solution of anomaly cancellation condition
found in \cite{bl961,nik961}. The
starting point is the following geometric data that defines
a rank $4$ stable vector bundle $V$ on a Calabi-Yau hypersurface 
$X$ in ${\Bbb P}(w_{1},\ldots , w_{5})$: 
\begin{eqnarray} 
0\rightarrow V \rightarrow \bigoplus_{a=1}^{5}
{\cal O}(q_{a})\stackrel{F_{a}}{\longrightarrow} 
{\cal O}(m)\rightarrow 0\label{2}
\end{eqnarray}
By setting $m=d$ and $\{q_{1},\ldots ,q_{5}\}=\{ w_{1},\ldots ,w_{5}\}$ the
equation (\ref{1}) is trivially satisfied. Assume that for one of the weights,
say $w_{5}$, we have $d/w_{5}\in 2{\Bbb Z}+1$. Replace $w_{5}$ by
$2w_{5}$ and define $w_{6}:=(m-w_{5})/2$. Furthermore, take
instead of $m$ the new integers $m_{1}:= m-w_{5}$ and $m_{2}:=
(m+3w_{5})/2$ into account. One can easily check that $\{ w_{1}, w_{2}, w_{3},
w_{4} ,w_{5} ; m \}$ and $\{ w_{1}, w_{2}, w_{3}, w_{4} ,2w_{5},w_{6} ;
m_{1},m_{2} \}$ satisfy 
\begin{eqnarray}
m^{2}-\sum^{5}_{i=1}w^{2}_{i}=m_{1}^{2}+m_{2}^{2}-
\sum^{4}_{i=1}w_{i}-(2w_{5})^{2}-w_{6}^{2}.
\end{eqnarray} 
In \cite{nik961,nik962} this equation has 
been interpreted as the anomaly cancellation condition for
the defining data of a $(0,2)$ Calabi-Yau 
$\sigma$ model with the same gauge bundle as in (\ref{2}), 
defined now on a Calabi-Yau complete
intersection in ${\Bbb P}(w_{1}, w_{2}, w_{3}, w_{4} ,2w_{5},w_{6})$.
Contrary to this interpretation we assume that these data describe
a $(0,2)$ supersymmetric $U(1)$ gauge theory  whose `Calabi-Yau phase'
is determined by the following exact sequence 
\begin{eqnarray}
0\rightarrow V \rightarrow \bigoplus_{i=1}^{4}
{\cal O}(w_{i})\oplus{\cal O}(2w_{5})\oplus{\cal O}(w_{6})
\stackrel{\mbox{\boldmath$\scriptstyle F$}}{\longrightarrow} 
{\cal O}(m_{1})\oplus{\cal O}(m_{2})\rightarrow 0\;\label{3}
\end{eqnarray}
on the same 
Calabi-Yau variety $X$ in ${\Bbb P}(w_{1},\ldots , w_{5})$ as before.
The choice of (\ref{3}) is technically
more appropriate because we have a good control on the reflexive polytopes
in four dimensions. However, it would be interesting to study  the
case of Calabi-Yau complete intersection and compare the results with
those of (\ref{3}).


\section{Intersection ring and Riemann-Roch theorem}

In this section we discuss some concepts of toric geometry which 
provide us with the main tools for the calculations of the next section. 
For the details
of the definitions and constructions used here 
we refer to the standard works \cite{DA78,OD88,FU93}.
In the first part of this section we use the homogeneous 
coordinate ring approach which is more appropriate for our
field theoretical considerations. 
The original motivation for its development  
was, however, the desire to have a construction of toric variety and related
objects similar to those of ${\Bbb P}^{n}$ in the classical 
algebraic geometry \cite{cox92}. \\

To begin with we first introduce some notation. Let $\bf N$ and 
${\bf M}=\mbox{Hom}({\bf N},{\Bbb Z})$ denote a dual pair of lattices of
rank $d$ and $\langle \cdot , \cdot \rangle$ be the canonical pairing on
${\bf M}\times {\bf N}$. Further, let ${\bf N}_{\Bbb R}={\bf N}
\otimes_{\Bbb Z}{\Bbb R}$ and  ${\bf M}_{\Bbb R}={\bf M}
\otimes_{\Bbb Z}{\Bbb R}$ be the $\Bbb R$-scalar extensions of
$\bf N$ and $\bf M$, respectively. $T={\bf N}\otimes_{\Bbb Z} 
{\Bbb C}^{*}=\mbox{Hom}_{\Bbb Z}({\bf M},{\Bbb C}^{*})$ is
the $d$ dimensional algebraic torus which acts on the 
toric variety ${\Bbb P}_{\tt \Sigma}$ defined by the  
(complete simplicial) fan $\Sigma$ in ${\bf N}_{\Bbb R}$ .
For a cone $\sigma \in \Sigma$ the dual cone, $\sigma^{\vee}$, is 
defined as usual by $\sigma^{\vee}=\{ m\in{\bf M}_{\Bbb R}\;|\; 
\langle m,n\rangle\geq 0 \;\; \mbox{for all} \;\; n\in \sigma \}$ 
and $\mbox{cosp}\sigma^{\vee}$ is the greatest
subspace of ${\bf M}_{\Bbb R}$ contained in $\sigma^{\vee}$.
The open affine variety in ${\Bbb P}_{\tt \Sigma}$ associated to $\sigma$ 
is denoted by $X_{\sigma^{\vee}}$.  
Let $\Sigma^{(k)}$ be the set of $k$ dimensional cones in $\Sigma$ .
By $e_{i}$ we denote the primitive lattice vectors on the 
one dimensional cones in $\Sigma^{(1)}=\{\rho_{1},
\ldots,\rho_{n}\}$. This set will play an important role in what follows.\\

Each one dimensional cone $\rho_{i}$ defines a $T$-invariant Weil divisor,
denoted by $D_{i}$, which is the closed subvariety 
$X_{\mbox{\scriptsize cosp}\rho_{i}^{\vee}}$ in $X_{\rho_{i}^{\vee}}$. 
This is indeed the
closed $T$-orbit associated to $\rho_{i}$ \cite{DA78}. The finitely generated
free abelian group $\bigoplus_{i=1}^{n}{\Bbb Z}\cdot D_{i}$ is the
group of $T$-invariant Weil divisors in ${\Bbb P}_{\tt \Sigma}$. Each
$m\in {\bf M}$ gives a character $\chi^{m}:T\to {\Bbb C}^{*}$, and hence
$\chi^{m}$ is a rational function on ${\Bbb P}_{\tt \Sigma}$. It defines
the Cartier divisor $\mbox{div}(\chi^{m})=\sum_{i=1}^{n}\langle m,e_{i}
\rangle \; D_{i}$ . In this way we obtain the map 
{\boldmath$\alpha$}
\begin{eqnarray}
\mbox{\boldmath$\alpha$}: {\bf M}\longrightarrow \bigoplus_{i=1}^{n}
{\Bbb Z}\cdot D_{i}\;\; ,
\;\; m\mapsto \sum_{i=1}^{n}\langle m,e_{i}\rangle \; D_{i}\;\label{4} .
\end{eqnarray}    
It follows from the completeness of the fan
$\Sigma$ that the map {\boldmath$\alpha$} is injective.
The cokernel of this map defines the Chow group $A_{d-1}
({\Bbb P}_{\tt \Sigma})$ which is a finitely generated abelian group
of rank $n-d$ . Therefore we have the following
exact sequence
\begin{eqnarray}
0\longrightarrow {\bf M}\stackrel{\mbox{\boldmath$\alpha$}}{\longrightarrow}
\bigoplus_{i=1}^{n}{\Bbb Z}\cdot D_{i}\stackrel{deg}{\longrightarrow}
A_{d-1}({\Bbb P}_{\tt \Sigma})\longrightarrow 0\;\label{5},
\end{eqnarray}
where $deg$ denotes the canonical projection. Now consider 
$G=\mbox{Hom}_{\Bbb Z}(A_{d-1}({\Bbb P}_{\tt \Sigma}),{\Bbb C}^{*})$
which is in general isomorphic to a product of 
$({\Bbb C}^{*})^{n-d}$ and a finite group.
By applying $\mbox{Hom}_{\Bbb Z}(\;\cdot \;, {\Bbb C}^{*})$ to (\ref{5}) we
get
\begin{eqnarray}
1\longrightarrow G \longrightarrow ({\Bbb C}^{*})^{n}\longrightarrow
T\longrightarrow 1\; ,
\end{eqnarray}
which defines the action of $G$ on ${\Bbb C}^{n}$ : $g\cdot 
{\mbox{\boldmath$x$}}
=(g( degD_{i})\; x_{i})$ for $g\in G$ and ${\mbox{\boldmath$x$}}
\in {\Bbb C}^{n}$.\\

Let $S={\Bbb C}[x_{1},\ldots,x_{n}]$ be the polynomial ring over $\Bbb C$
with variables $x_{1},\ldots,x_{n}$, where  $x_{i}$ correspond to the
one dimensional cones $\rho_{i}$ in $\Sigma$ . This ring
is graded in a natural way by $deg(x_{i}):=degD_{i}$ .
Then, let $B$ be the monomial ideal in $S$ generated by $\prod_{\rho_{i}
\not\subset \sigma }x_{i}$ for all $\sigma\in\Sigma$ . The ring $S$
defines the $n$-dimensional affine space ${\Bbb A}^{n}= 
\mbox{Spec}(S)$. The ideal $B$ gives the variety 
\begin{eqnarray}
{\bf Z}_{\tt \Sigma}={\bf V}(B)
\end{eqnarray}
which is denoted as the exceptional set. 
Removing the exceptional set ${\bf Z}_{\tt \Sigma}$ we obtain
the Zariski open set
\begin{eqnarray}
{\bf U}_{\tt \Sigma}= {\Bbb A}^{n}\setminus{\bf Z}_{\tt \Sigma}\; , 
\end{eqnarray}
which is invariant under the action of $G$.
For the case of a complete simplicial fan  the geometric
quotient of ${\bf U}_{\tt \Sigma}$ by $G$ exists and gives rise to 
${\Bbb P}_{\tt \Sigma}$ \cite{cox92}!\\

Having reviewed these preliminary concepts 
we now describe the process of resolving
the singularities. At first we have to resolve
the singularities of the base variety. We begin with
a reflexive polytope $\Delta$ in $\bf N$ which corresponds to the weighted
projective space ${\Bbb P}(w_{1},\ldots,w_{5})$. Let $\Sigma$ be the 
fan in ${\bf N}_{\Bbb R}$ associated to $\Delta$. 
The toric variety ${\Bbb P}_{\tt \Sigma}$ then has an ample 
anticanonical sheaf, whose generic section realizes our (canonical) 
Calabi-Yau variety.\\ 

Taking a maximal triangulation of $\Delta$ leads in
our case , i.e. $d=4$, to a fully resolved Calabi-Yau 
variety $\widetilde X$ \cite{ba94}. 
A maximal triangulation of $\Delta$ amounts above all to adding new one
dimensional cones to $\Sigma^{(1)}$ which are associated to the 
points on the faces of $\Delta$. In the context of gauged linear sigma
models these correspond to the additional chiral scalar superfields.
As mentioned in the previous section we also need to determine the charges
of the fields with respect to the full gauge group. Translated into the
geometric language this means that we have to determine the grading of
the variables in the ring S. Using (\ref{4}) this can be done by 
solving 
\begin{eqnarray}
\sum_{i=1}^{n}w_{i}^{(k)}\mbox{\boldmath$\alpha$}_{ij}=0 \;\;\;
\mbox{for all}\;\;\;k=1,\ldots,N\;\mbox{and}\; j=1,\ldots,d\;  
\end{eqnarray}
which gives the charges $\mbox{\boldmath $w$}_{i}$ of $x_{i}$ : 
$\mbox{\boldmath $w$}_{i}=deg(x_{i})=
(w_{i}^{(1)},\ldots,w_{i}^{(\tt N)})$.
Note that the desingularization of the base variety simultaneously resolves
the tangent sheaf to which the right-moving fermions couple. 
Therefore, these fermions have the same charges as their superpartners.\\

What about the left-moving fermions? The geometric data of the gauge
bundle $V$ in a $(0,2)$ model are the additional degrees of freedom 
which we still have to deal with. After resolving the singularities
of the base variety we pull the exact sequence (\ref{3}) back to the
desingularized base variety. Because this process only preserves 
the right exactness of this sequence we are forced to `modify' it
in oder to get the exact sequence
\begin{eqnarray}
0\rightarrow \widetilde{V} \rightarrow \bigoplus_{i=1}^{6}
{\cal O}(\mbox{\boldmath$q$}_{i})
\stackrel{\widetilde F}{\longrightarrow} 
\bigoplus_{j=1}^{2}{\cal O}(\mbox{\boldmath$p$}_{j})\rightarrow 0\;\label{6} ,
\end{eqnarray}       
where $\mbox{\boldmath$q$}_{i}=(q_{i}^{(1)},\ldots,q_{i}^{(\tt N)})$ 
and $\mbox{\boldmath$p$}_{j}=(p_{j}^{(1)},\ldots,p_{j}^{(\tt N)})$ denote
the degrees (or charges) of the pulled back divisors in the
desingularized base variety, which were originally associated to rank
one sheaves in (\ref{3}). 
Following \cite{di96} we impose the same conditions as before on this data 
guaranteeing the existence of spinors and the cancellation of 
the gauge anomaly. The conditions $c_{1}({\widetilde{T_{\scriptstyle X}}})= 
c_{1}({\widetilde V})=0$ result in
\begin{eqnarray}
\mbox{\boldmath$d$}=\sum_{i=1}^{5}\mbox{\boldmath$w$}_{i} 
\;\;\;\; \mbox{and}  \;\;\;\;
\mbox{\boldmath$p$}_{1}+\mbox{\boldmath$p$}_{2}=\sum_{i=1}^{6}
\mbox{\boldmath$q$}_{i}\label{3a}\; ,
\end{eqnarray}
where $\widetilde{T_{\scriptstyle X}}$ denotes the resolved tangent sheaf
and $\mbox{\boldmath$d$}=(d^{(1)},\ldots,d^{(\tt N)})$ is 
the degree of the pulled back divisor corresponding to the last term in 
(\ref{1a}). The anomaly cancellation condition leads to the 
following Diophantine equations
\begin{eqnarray}
d^{(l)}d^{(k)}-\sum_{i=1}^{5}w_{i}^{(l)}w_{i}^{(k)}=
\sum_{j=1}^{2}p_{j}^{(l)}p_{j}^{(k)}-\sum_{i=1}^{6}q_{i}^{(l)}q_{i}^{(k)}
\hspace{8mm}\mbox{for}\;\;\;\;l,k=1,\ldots,N\label{3b}\; .
\end{eqnarray}
Each solution of these equations gives a possible gauge bundle data for
the desingularized theory. One should be careful about the exactness of
(\ref{6}). It may happen that one can not choose polynomials
${\widetilde F}$'s such
that not all of them vanish simultaneously on the base variety. If
this is the case, then one has to deal with a sheaf which is no longer
locally free (cf. \cite{di96} for details). The examples that will be 
considered here avoid this problem.\\

Now we come to the discussion of intersection ring. Let 
\begin{eqnarray}
A_{*}({\Bbb P}_{\tt\Sigma})=\bigoplus_{k}A_{k}({\Bbb P}_{\tt\Sigma}) 
\end{eqnarray} 
be the Chow ring of ${\Bbb P}_{\tt\Sigma}$, where 
$A_{k}({\Bbb P}_{\tt\Sigma})$ are the finitely 
generated abelian groups of $k$-cycles\footnote{These are
the closed $T$-orbits associated to the elements of $\Sigma^{(d-k)}$.}
in ${\Bbb P}_{\tt\Sigma}$ up to rational
equivalence. The multiplicative structure of
$A_{*}({\Bbb P}_{\tt\Sigma})$ is given by the intersection of cycles. 
It is determined by the combinatorial data encoded
in the fan $\Sigma$ . If $\sigma_{1}$ and $\sigma_{2}$ are two cones in
$\Sigma$ that are faces of at least one other cone $\tau$ in the fan, 
then their corresponding cycles do intersect otherwise 
the intersection is empty. The intersection cycle in the former case
is the one that is associated to $\tau$. It is conveniet to consider
$A^{k}({\Bbb P}_{\tt\Sigma}):=A_{d-k}({\Bbb P}_{\tt\Sigma})$ because
by intersecting cycles the codimensions add up. Therefore, we obtain
the graded commutative ring
\begin{eqnarray}
A^{*}({\Bbb P}_{\tt\Sigma})=\bigoplus_{k}A^{k}({\Bbb P}_{\tt\Sigma}) 
\end{eqnarray}
which is called the intersection ring of ${\Bbb P}_{\tt\Sigma}$.
Let ${\Bbb Q}[z_{1},\ldots,z_{n}]$ be the polynomial ring in 
variables $z_{1},\ldots,z_{n}$ over $\Bbb Q$, where
$z_{i}$ correspond to the one dimensional cones $\rho_{i}$ in the fan 
$\Sigma$. Further, let $I=\langle \;\sum_{i=1}^{n}\langle m,e_{i}\rangle 
\; z_{i} : m\in {\bf M}\;\rangle$ and $J=\langle \;\prod_{\rho_{i}
\in{\cal P}}z_{i}  : \mbox{for all}\; {\cal P}\;\rangle$ 
(is called Stanley--Reisner ideal) be
ideals in ${\Bbb Q}[z_{1},\ldots,z_{n}]$, where $\cal P$ stands for
a primitive collection. It is a subset $\{\rho_{i_{1}},\ldots,
\rho_{i_{k}}\}$ of $\Sigma^{(1)}$ which does not generate  
a $k$-dimensional cone, whereas any proper subset of it 
generates a cone in $\Sigma$. It can be shown that for a complete
simplicial toric variety ${\Bbb P}_{\tt\Sigma}$  one has the
follwoing isomorphisms for the intersection
ring $A^{*}({\Bbb P}_{\tt\Sigma})_{\Bbb Q}\;(= A^{*}({\Bbb P}_{\tt\Sigma})
\otimes_{\Bbb Z}{\Bbb Q})$ 
\begin{eqnarray}
A^{*}({\Bbb P}_{\tt\Sigma})_{\Bbb Q}\;\simeq\;{\Bbb Q}[z_{1},\ldots,z_{n}]/
(I+J)\;\simeq\; H^{*}({\Bbb P}_{\tt\Sigma},{\Bbb Q})\;,
\end{eqnarray}  
where the isomorphism in the direction of the rational cohomology ring 
doubles the degree. We are actually interested in the intersection ring
of the desingularized Calabi-Yau variety $\widetilde X$ 
which, as pointed out above, is a generic section of the 
anticanonical sheaf on ${\Bbb P}_{\tt\Sigma}$. The ring 
$A^{*}({\widetilde X})_{\Bbb Q}$
is isomorphic to the quotient of $A^{*}({\Bbb P}_{\tt\Sigma})_{\Bbb Q}$
by the annihilator of the canonical divisor \cite{ba93}, i.e.
\begin{eqnarray}
A^{*}({\widetilde X})_{\Bbb Q}\;\simeq\;A^{*}({\Bbb P}_{\tt\Sigma})
_{\Bbb Q}/\mbox{ann}(z_{1}+\ldots+z_{n})\;.
\end{eqnarray}

Let $\cal F$ be a coherent sheaf on the (complete) variety {\boldmath$X$}.
As we know, the Euler characteristic of $\cal F$, $\chi(\mbox{\boldmath$X$},
{\cal F})$, is defined by
\begin{eqnarray}
\chi({\cal F})=\chi(\mbox{\boldmath$X$},{\cal F})=\sum_{p\geq 0}
(-1)^{p}\; H^{p}(\mbox{\boldmath$X$},{\cal F})\;.
\end{eqnarray}
The main property of the Euler characteristic is its additivity. It
means that for an exact sequence of coherent sheaves $0\to{\cal E}\to
{\cal F}\to{\cal G}\to 0$ it holds that $\chi({\cal F})=\chi({\cal E})+
\chi({\cal G})$.\\

The Riemann-Roch theorem allows us to express the
Euler characteristic of a coherent sheaf $\cal F$ on the (smooth)
variety {\boldmath$X$} in terms of the intersection of algebraic 
cycles in {\boldmath$X$}. We now define the intersection form
$(\cdot,\cdot)$ on $A^{*}(\mbox{\boldmath$X$})$ 
which we need in the statement of
the Riemann-Roch theorem. It is a map $A^{*}(\mbox{\boldmath$X$})\times
A^{*}(\mbox{\boldmath$X$})\to {\Bbb Q}$ that is the composition of
the multiplication in $A^{*}(\mbox{\boldmath$X$})$ and the linear
functional $\mbox{deg}:A^{*}(\mbox{\boldmath$X$})\to{\Bbb Q}$ defined
as follows. deg associates to each cycle $\eta\in A^{*}
(\mbox{\boldmath$X$})$ the degree of its $0$-cycle part $\eta_{0}=
\sum_{i}n_{i}[P_{i}]$ : $\mbox{deg}(\eta_{0})=\sum_{i}n_{i}$ . For a
smooth projective variety {\boldmath$X$} the Riemann-Roch theorem reads
\begin{eqnarray}
\chi({\cal F})=( \mbox{ch}({\cal F}), \mbox{Td} (\mbox{\boldmath$X$}))\;,
\end{eqnarray}    
where $\mbox{ch}({\cal F})$ is the Chern character of $\cal F$ and 
$\mbox{Td}(\mbox{\boldmath$X$})=\mbox{td}
(T_{\mbox{\boldmath$\scriptstyle X$}})$ is the Todd class of the tangent
sheaf. $\mbox{ch}({\cal F})$ and $\mbox{Td}(\mbox{\boldmath$X$})$ are
elements of $A^{*}(\mbox{\boldmath$X$})$! We recall that the Chern 
character of a coherent sheaf $\cal F$ is defined through its locally
free resolution. For a locally free sheaf $\cal E$ of rank $r$ we have:
\begin{eqnarray}
\mbox{ch}({\cal E}) & = & r+c_{1}+\frac{1}{2}(c_{1}^{2}-2c_{2})+\frac{1}{6}
                          (c_{1}^{3}-3c_{1}c_{2}+3c_{3})\nonumber\\
                    &   & +\frac{1}{24}(c_{1}^{4}-4c_{1}^{2}c_{2}
                          +4c_{1}c_{3}+2c_{2}^{2}-4c_{4})+\dots\label{7}\\
\mbox{td}({\cal E}) & = & 1+\frac{1}{2}c_{1}+\frac{1}{12}(c_{1}^{2}+c_{2})
                           +\frac{1}{24}c_{1}c_{2}\nonumber\\
                    &   & -\frac{1}{720}(c_{1}^{4}-4c_{1}^{2}c_{2}-3c_{2}^{2}
                           -c_{1}c_{3}+c_{4})+\ldots\; \label{8},
\end{eqnarray}
where $c_{i}=c_{i}({\cal E})$ is the $i$-th Chern class of $\cal E$.


\section{Examples}
In this section we discuss a few examples of the class of models 
given by the geometric data of (\ref{3}) on a Calabi-Yau variety in
the weighted projevtive space ${\Bbb P}(w_{1},\ldots,w_{5})$.\\

\subsubsection*{Example 1 : ${\Bbb P}(1,1,1,3,3)$}

The gauge bundle $V$ is given by 
\begin{eqnarray}
0\to V\to {\cal O}(1)^{\oplus 3}\oplus{\cal O}(3)
\oplus{\cal O}(6)\oplus{\cal O}(3)\to {\cal O}(6)\oplus{\cal O}(9)
\to 0\nonumber\;.
\end{eqnarray}
The reflexive polytope $\Delta$ corresponding to ${\Bbb P}(1,1,1,3,3)$
is defined by the vertices 
\begin{eqnarray}
&e_{1}=(1,0,0,0), e_{2}=(0,1,0,0), e_{3}=(0,0,1,0)&\nonumber\\
&e_{4}=(0,0,0,1), e_{5}=(-1,-1,-3,-3)& \nonumber
\end{eqnarray}
with respect to canonical basis in the lattice $\bf N$. Apart from the
unique inner point, $\Delta$ still has one additional point $e_{6}$ which
lies on the codimension 2 face of $\Delta$ generated as the convex
hull of the points $e_{1},e_{2}$ and $e_{5}$ : $e_{6}=(0,0,-1,-1)$. 
Therefore, the desingularization in
this case gives rise to an extra $U(1)$ factor. We now proceed to find 
the charges of the fields. As before  we denote by $x_{i}$ and $D_{i}$ 
the variables in $S$ and the divisors associated to $e_{i}$.
In the canonical basis $\{u_{i}\}_{i=1}^{4}$ of $\bf M$, the
map {\boldmath$\alpha$} is represented by
\begin{eqnarray}
\left(
\begin{array}{cccc}
1 & 0 & 0 & 0\\
0 & 1 & 0 & 0\\
0 & 0 & 1 & 0\\
0 & 0 & 0 & 1\\
-1&-1 &-3 &-3\\
0 & 0 & -1&-1
\end{array}
\right)\nonumber\; ,
\end{eqnarray}  
which yields 
\begin{center}
\begin{tabular}{||c||c|c|c|c|c||c\TVR74||} \hline\hline
field & $x_{1}$ & $x_{2}$ & $x_{3}$ & $x_{4}$ & $x_{5}$ & $x_{6}$\\ \hline
charge& $(1,0)$ & $(1,0)$ & $(3,1)$ & $(3,1)$ & $(1,0)$ & $(0,1)$\\\hline\hline
\end{tabular}
\end{center}    
Using this table we determine in the next step the data of
the resolved gauge bundle $\widetilde V$. From (\ref{3a}) and (\ref{3b}) 
we obtain
\begin{eqnarray}
& & q_{1}+ q_{2}+ q_{3}+ q_{4}+ q_{5}+ q_{6}-p_{1}-p_{2}=0\;,\nonumber\\
& & q_{1}+ q_{2}+ q_{3}+ 3q_{4}+6q_{5}+ 3q_{6}-6p_{1}-9p_{2}=-12\;,\nonumber\\
& & q_{1}^{2}+ q_{2}^{2}+ q_{3}^{2}+ q_{4}^{2}+ q_{5}^{2}+ q_{6}^{2}
-p_{1}^{2}-p_{2}^{2}=-2\;,\nonumber
\end{eqnarray} 
where we have dropped the index $(2)$ on $q$'s and $p$'s. Here is a set of
solutions of the above Diophantine equations
\begin{center}
\begin{tabular}{||c|c|c|c|c|c||c|c||} \hline\hline
$q_{1}$ &$q_{2}$ &$q_{3}$ &$q_{4}$ &$q_{5}$ &$q_{6}$ &$p_{1}$ &$p_{2}$\\\hline
  $0$  &  $-1$  &  $1$  &  $1$  &  $4$  &  $0$  &  $2$  &  $3$  \\\hline
  $0$  &  $0$  &  $0$  &  $1$  &  $0$  &  $1$  &  $0$  &  $2$  \\\hline
  $1$  &  $1$  &  $0$  &  $0$  &  $0$  &  $0$  &  $2$  &  $0$  \\\hline
  $0$  &  $0$  &  $0$  &  $1$  &  $1$  &  $1$  &  $1$  &  $2$  \\\hline\hline
\end{tabular}
\end{center}   
which result in 
\begin{center}
\begin{tabular}{||c||c\TVR74||} \hline\hline
$({\bf a})$& $ 0\to{\widetilde V}\to {\cal O}(1,0)\oplus{\cal O}(1,-1)
                                          \oplus{\cal O}(1,1)\oplus{\cal O}
                                          (6,4)$\\
           &                            $ \oplus{\cal O}(3,1)\oplus{\cal O}
                                         (3,0)\to{\cal O}(6,2)\oplus
                                         {\cal O}(9,3)\to 0$\\\hline

$({\bf b})$& $ 0\to{\widetilde V}\to {\cal O}(1,0)\oplus{\cal O}(1,0)
                                         \oplus{\cal O}(1,0)\oplus{\cal O}
                                         (3,1)$\\
           &                            $\oplus{\cal O}(6,0)\oplus
                                          {\cal O}(3,1)\to{\cal O}(6,0)
                                          \oplus{\cal O}(9,2)\to 0$\\\hline

$({\bf c})$& $ 0\to{\widetilde V}\to {\cal O}(1,1)\oplus{\cal O}(1,1)
                                          \oplus{\cal O}(1,0)\oplus{\cal O}
                                          (3,0)$\\
           &                            $\oplus{\cal O}(6,0)\oplus
                                          {\cal O}(3,0)\to{\cal O}(6,2)
                                          \oplus{\cal O}(9,0)\to 0$\\\hline
$({\bf d})$& $ 0\to{\widetilde V}\to {\cal O}(1,0)\oplus{\cal O}(1,0)
                                          \oplus{\cal O}(1,0)
                                          \oplus{\cal O}(3,1)$\\
           &                           $\oplus{\cal O}(6,1)\oplus{\cal O}
                                         (3,1)\to{\cal O}(6,1)\oplus
                                         {\cal O}(9,2)\to 0$\\\hline\hline

\end{tabular}
\end{center}  
We now consider a fan $\Sigma$ corresponding to the maximal 
triangulation of $\Delta$ given by the `big cones'
\begin{eqnarray}
\langle e_{1}e_{2}e_{3}e_{4}\rangle,\langle e_{1}e_{2}e_{3}e_{6}\rangle,
\langle e_{1}e_{2}e_{4}e_{6}\rangle,\nonumber\\
\langle e_{1}e_{3}e_{4}e_{5}\rangle,\langle e_{1}e_{3}e_{5}e_{6}\rangle,
\langle e_{1}e_{4}e_{5}e_{6}\rangle,\nonumber\\  
\langle e_{2}e_{3}e_{4}e_{5}\rangle,\langle e_{2}e_{3}e_{5}e_{6}\rangle,
\langle e_{2}e_{4}e_{5}e_{6}\rangle\nonumber,  
\end{eqnarray}
where $\langle e_{i}e_{j}e_{k}e_{l}\rangle$ denotes the cone generated
by $e_{i},e_{j},e_{k}$ and $e_{l}$. The primitive collections of $\Sigma$
are $\{ e_{3},e_{4},e_{6}\}$ and $\{e_{1},e_{2},e_{5}\}$. With these
combinatorial data at hand we can write down the ideals 
$I$ and $J$:
\begin{eqnarray}
I=\langle z_{1}-z_{5},z_{2}-z_{5},z_{3}-3z_{5}-z_{6},
z_{4}-3z_{5}-z_{6}\rangle\;\;,\;\;J=\langle z_{1}z_{2}z_{5},
z_{3}z_{4}z_{6}\rangle\nonumber\;.
\end{eqnarray}
Because we are going to make calculations in a polynomial ring, it is
convenient to use the Gr\"{o}bner basis method \cite{ei95,ad94,cox921}.
A Gr\"{o}bner basis of $I+J$ with respect to the lex order
$z_{1}>\ldots>z_{6}$ is given by
\begin{eqnarray}
I+J&=\langle & z_{1}-z_{5}\;,\;z_{2}-z_{5}\;,\;z_{3}-3z_{5}-z_{6}\;,\;
               z_{4}-3z_{5}-z_{6},\nonumber\\
   &         & z_{5}^{3} \;,\; 9z_{5}^{2}z_{6}+6z_{5}z_{6}^{2}+z_{6}^{3}\;,\;
               9z_{5}z_{6}^{3}+2z_{6}^{4}\;,\; z_{6}^{5}\; \rangle\nonumber\;.
\end{eqnarray}
Let $K$ be the ideal in the polynomial 
ring ${\Bbb Q}[z_{1},\ldots,z_{6}]$ generated by $z_{1}+\ldots+z_{6}$ which
is a representative of the canonical class in the intersection ring
$A^{*}({\Bbb P}_{\tt\Sigma})$. Then the annihilator of $z_{1}+\ldots+z_{6}$
in $A^{*}({\Bbb P}_{\tt\Sigma})$ is given by
\begin{eqnarray}
\mbox{ann}(z_{1}+\ldots+z_{6})=(I+J):K\nonumber\;,
\end{eqnarray}
where `$\,:\,$' denotes the quotient of ideals. 
$\mbox{ann}(z_{1}+\ldots+z_{6})$ in this
example is calculated to
\begin{eqnarray}
\mbox{ann}(z_{1}+\ldots+z_{6})&=\langle& z_{1}+z_{2}+z_{3}+z_{4}-8z_{5}-2z_{6}\;,\;
                                  z_{2}-z_{5},\nonumber\\ 
                       &        & z_{3}-3z_{5}-z_{6}\;,\;z_{4}-3z_{5}-z_{6}\;,
                                  \;z_{5}^{3}\;,\;3z_{5}z_{6}+z_{6}^{2}\;,
                                  \;z_{6}^{4}\;\rangle\nonumber\;.
\end{eqnarray}
A look at (\ref{7}) and (\ref{8}) shows that in the product of
$\mbox{ch}({\widetilde V})$ and $\mbox{Td}({\widetilde X})$ in the 
intersection ring $A^{*}({\widetilde X})$ the only $0$-cycle part
is given by $1/2\; c_{3}({\widetilde V})$. Applying the degree functional 
to this term 
gives us the Euler characteristic of the respective gauge bundle. One
should be careful about the normalization of the product of cycles.
For the big cones of $\Sigma$ the normalization is fixed by 
\begin{eqnarray}
\langle D_{i_{1}} \ldots  D_{i_{4}}\rangle=
\frac{1}{\mbox{mult}(e_{i_{1}},\ldots,e_{i_{4}})}\label{9}\;\;,
\end{eqnarray}
where $\mbox{mult}(e_{i_{1}},\ldots,e_{i_{4}})$ denotes the index in
$\bf N$ of the lattice spanned by these vectors. 
Multiplying the top terms in the intersection
ring of the desingularized Calabi-Yau variety by the representative of
the canonical divisor and using (\ref{9}) together with the `algebraic
moving lemma' \cite{DA78,FU93} yields the normalization in
$A^{*}({\widetilde X})$.\\

All big cones in this example have volume one. Therefore, 
$\langle D_{i_{1}} \ldots D_{i_{4}}\rangle = 1$ for all 
big cones in $\Sigma$. As we will see 
below, the third Chern class of the gauge bundle is represented by a
degree three monomial in $z_{6}$. Its normalization in 
$A^{*}({\widetilde X})$ is given by $\langle z_{6}^{3}\rangle = 27$.
We have summarized the result of the calculations for the resolved bundles
found above in the following table
\begin{center}
\begin{tabular}{||c|c||c\TVR74||} \hline\hline
$\widetilde V$ & $c_{3}({\widetilde V})$ & $\chi({\widetilde V})$\\\hline\hline
$({\bf a})$    & $-\frac{20}{3}z_{6}^{3}$ & $\bf -90$ \\\hline
$({\bf b})$    & $-8z_{6}^{3}$            & $\bf -108$\\\hline
$({\bf c})$    & $-\frac{22}{9}z_{6}^{3}$ & $\bf -33$\\\hline 
$({\bf d})$    & $-8z_{6}^{3}$            & $\bf -108$\\\hline\hline
\end{tabular}
\end{center}

\subsubsection*{Example 2 : ${\Bbb P}(1,1,2,2,3)$}

The gauge bundle $V$ is given by 
\begin{eqnarray}
0\to V\to {\cal O}(1)^{\oplus 2}\oplus{\cal O}(2)^{\oplus 2}
\oplus{\cal O}(6)\oplus{\cal O}(3)\to {\cal O}(6)\oplus{\cal O}(9)
\to 0\nonumber\;.
\end{eqnarray}
The reflexive polytope $\Delta$ corresponding to ${\Bbb P}(1,1,2,2,3)$
is defined by the vertices 
\begin{eqnarray}
&e_{1}=(1,0,0,0), e_{2}=(0,1,0,0), e_{3}=(0,0,1,0)&\nonumber\\
&e_{4}=(0,0,0,1), e_{5}=(-1,-2,-2,-3), e_{6}=(0,-1,-1,-1)\nonumber\;.&
\end{eqnarray}
This exhausts the set of boundary points of $\Delta$. The desingularization
gives rise as before to an extra $U(1)$ factor. The map 
{\boldmath$\alpha$} is represented by
\begin{eqnarray}
\left(
\begin{array}{cccc}
1 & 0 & 0 & 0\\
0 & 1 & 0 & 0\\
0 & 0 & 1 & 0\\
0 & 0 & 0 & 1\\
-1&-2 &-2 &-3\\
0 & -1 & -1&-1
\end{array}
\right)\nonumber\; ,
\end{eqnarray}  
which yields 
\begin{center}
\begin{tabular}{||c||c|c|c|c|c||c\TVR74||} \hline\hline
field & $x_{1}$ & $x_{2}$ & $x_{3}$ & $x_{4}$ & $x_{5}$ & $x_{6}$\\ \hline
charge& $(1,0)$ & $(2,1)$ & $(2,1)$ & $(3,1)$ & $(1,0)$ & $(0,1)$\\\hline\hline
\end{tabular}
\end{center}     
Using this table we obtain the following equations for the data of
the resolved gauge bundle $\widetilde V$
\begin{eqnarray}
& & q_{1}+ q_{2}+ q_{3}+ q_{4}+ q_{5}+ q_{6}-p_{1}-p_{2}=0\;,\nonumber\\
& & q_{1}+ q_{2}+ 2q_{3}+ 2q_{4}+6q_{5}+ 3q_{6}-6p_{1}-9p_{2}=-20\;,\nonumber\\
& & q_{1}^{2}+ q_{2}^{2}+ q_{3}^{2}+ q_{4}^{2}+ q_{5}^{2}+ q_{6}^{2}
-p_{1}^{2}-p_{2}^{2}=-6\;.\nonumber
\end{eqnarray} 
Two solutions of these equations are
\begin{center}
\begin{tabular}{||c|c|c|c|c|c||c|c||} \hline\hline
$q_{1}$ &$q_{2}$ &$q_{3}$ &$q_{4}$ &$q_{5}$ &$q_{6}$ &$p_{1}$ &$p_{2}$\\\hline
  $1$  &  $1$  &  $1$  &  $0$  &  $2$  &  $0$  &  $3$  &  $2$  \\\hline
  $2$  &  $0$  &  $1$  &  $0$  &  $3$  &  $0$  &  $4$  &  $2$  \\\hline\hline
\end{tabular}
\end{center}   
which lead to 
\begin{center}
\begin{tabular}{||c||c\TVR74||} \hline\hline
$({\bf a})$& $ 0\to{\widetilde V}\to {\cal O}(1,1)\oplus{\cal O}(1,1)
                                          \oplus{\cal O}(2,1)\oplus{\cal O}
                                          (2,0)$\\
           &                            $ \oplus{\cal O}(6,2)\oplus{\cal O}
                                         (3,0)\to{\cal O}(6,3)\oplus
                                         {\cal O}(9,2)\to 0$\\\hline

$({\bf b})$& $ 0\to{\widetilde V}\to {\cal O}(1,2)\oplus{\cal O}(1,0)
                                         \oplus{\cal O}(2,1)\oplus{\cal O}
                                         (2,0)$\\
           &                            $\oplus{\cal O}(6,3)\oplus
                                          {\cal O}(3,0)\to{\cal O}(6,4)
                                         \oplus{\cal O}(9,2)\to 0$\\\hline\hline
\end{tabular}
\end{center}  
The big cones of the fan $\Sigma$ corresponding to the maximal triangulation
of $\Delta$ are
\begin{eqnarray}
\langle e_{1}e_{2}e_{3}e_{4}\rangle,\langle e_{1}e_{2}e_{3}e_{5}\rangle,
\langle e_{1}e_{2}e_{4}e_{6}\rangle,\nonumber\\
\langle e_{1}e_{2}e_{5}e_{6}\rangle,\langle e_{1}e_{3}e_{4}e_{6}\rangle,
\langle e_{1}e_{3}e_{5}e_{6}\rangle,\nonumber\\  
\langle e_{2}e_{3}e_{4}e_{5}\rangle,\langle e_{2}e_{4}e_{5}e_{6}\rangle,
\langle e_{3}e_{4}e_{5}e_{6}\rangle\nonumber\;.  
\end{eqnarray}
The primitive collections of $\Sigma$ are $\{ e_{2},e_{3},e_{6}\}$ 
and $\{e_{1},e_{4},e_{5}\}$. Using these combinatorial data we find
$I=\langle\; z_{1}-z_{5}\;,\;z_{2}-2z_{5}-z_{6}\;,\;
z_{3}-2z_{5}-z_{6}\;,\;z_{4}-3z_{5}-z_{6}\;\rangle$ and 
$J=\langle\; z_{1}z_{4}z_{5}\;,\;z_{2}z_{3}z_{6}\;\rangle$. With
respect to the lex order $z_{1}>\ldots>z_{6}$ the Gr\"{o}bner bases
of $I+J$ and $\mbox{ann}(z_{1}+\ldots+z_{6})$ are given by
\begin{eqnarray}
I+J & = \langle & z_{1}-z_{5}\;,\;z_{2}-2z_{5}-z_{6}\;,\;
                  z_{3}-2z_{5}-z_{6}\;,\;z_{4}-3z_{5}-z_{6}\;,\;\nonumber\\
    &           & 3z_{5}^{3}+z_{5}^{2}z_{6}\;,\;4z_{5}^{2}z_{6}+4z_{5}
                  z_{6}^{2}+z_{6}^{3}\;,\;5z_{5}z_{6}^{3}+2z_{6}^{4}\;,\;
                  z_{6}^{5}\;\rangle \nonumber\\ 
\mbox{ann}(z_{1}+\ldots+z_{6})&=\langle& z_{1}+z_{2}+z_{3}+z_{4}-8z_{5}-3z_{6}\;,\;
                                  z_{2}-2z_{5}-z_{6}\;,\;
                                  z_{3}-2z_{5}-z_{6}\;,\;\nonumber\\
                       &        & z_{4} -3z_{5}-z_{6}\;,\;z_{5}^{2}+7z_{5}
                                  z_{6}+4z_{6}^{2}\;,\;8z_{5}z_{6}^{2}+
                                  5z_{6}^{3}\;,\;z_{6}^{4}\;\rangle\nonumber\;. 
\end{eqnarray} 
The normalization in this case is as follows: $\langle D_{1} D_{2}
D_{3} D_{5}\rangle = \frac{1}{3}$ and all other big cones
have unit volume. This leads to the normalization 
$\langle z_{6}^{3}\rangle =8$ in the intersection ring 
$A^{*}({\widetilde X})$. Therefore, we obtain
\begin{center}
\begin{tabular}{||c|c||c\TVR74||} \hline\hline
$\widetilde V$ & $c_{3}({\widetilde V})$ & $\chi({\widetilde V})$\\\hline\hline
$({\bf a})$    & $-\frac{51}{4}z_{6}^{3}$ & $\bf -51$ \\\hline
$({\bf b})$    & $-\frac{21}{2}z_{6}^{3}$ & $\bf -42$\\\hline\hline
\end{tabular}
\end{center}

\subsubsection*{Example 3 : ${\Bbb P}(1,2,2,3,4)$}

The gauge bundle $V$ is given by 
\begin{eqnarray}
0\to V\to {\cal O}(1)\oplus{\cal O}(2)^{\oplus 2}
\oplus{\cal O}(3)\oplus{\cal O}(8)\oplus {\cal O}(4)\to{\cal O}(8)
\oplus{\cal O}(12)\to 0\nonumber\;.
\end{eqnarray}
The reflexive polytope $\Delta$ corresponding to ${\Bbb P}(1,2,2,3,4)$
is defined by the vertices 
\begin{eqnarray}
&e_{1}=(1,0,0,0), e_{2}=(0,1,0,0), e_{3}=(0,0,1,0)&\nonumber\\
&e_{4}=(0,0,0,1), e_{5}=(-2,-2,-3,-4)\;.& \nonumber
\end{eqnarray}
$\Delta$ still has one other boundary point $e_{6}=(-1,-1,-1,-2)$ which 
lies on the codimension three face generated by $e_{3}$ and $e_{5}$. The 
desingularization gives rise as before to an extra $U(1)$ factor. The map 
{\boldmath$\alpha$} is represented by
\begin{eqnarray}
\left(
\begin{array}{cccc}
1 & 0 & 0 & 0\\
0 & 1 & 0 & 0\\
0 & 0 & 1 & 0\\
0 & 0 & 0 & 1\\
-2&-2 &-3 &-4\\
-1& -1 & -1&-2
\end{array}
\right)\nonumber\; ,
\end{eqnarray}  
which yields 
\begin{center}
\begin{tabular}{||c||c|c|c|c|c||c\TVR74||} \hline\hline
field & $x_{1}$ & $x_{2}$ & $x_{3}$ & $x_{4}$ & $x_{5}$ & $x_{6}$\\ \hline
charge& $(2,1)$ & $(2,1)$ & $(3,1)$ & $(4,2)$ & $(1,0)$ & $(0,1)$\\\hline\hline
\end{tabular}
\end{center}     
Therefore, we obtain the following equations for the data of
the resolved gauge bundle $\widetilde V$
\begin{eqnarray}
& & q_{1}+ q_{2}+ q_{3}+ q_{4}+ q_{5}+ q_{6}-p_{1}-p_{2}=0\;,\nonumber\\
& & q_{1}+ 2q_{2}+2q_{3}+3q_{4}+8q_{5}+ 4q_{6}-8p_{1}-12p_{2}=-45\;,\nonumber\\
& & q_{1}^{2}+ q_{2}^{2}+ q_{3}^{2}+ q_{4}^{2}+ q_{5}^{2}+ q_{6}^{2}
-p_{1}^{2}-p_{2}^{2}=-18\;.\nonumber
\end{eqnarray} 
Two solutions of these equations are
\begin{center}
\begin{tabular}{||c|c|c|c|c|c||c|c||} \hline\hline
$q_{1}$ &$q_{2}$ &$q_{3}$ &$q_{4}$ &$q_{5}$ &$q_{6}$ &$p_{1}$ &$p_{2}$\\\hline
  $1$  &  $1$  &  $1$  &  $2$  &  $0$  &  $2$  &  $5$  &  $2$  \\\hline
  $2$  &  $0$  &  $2$  &  $3$  &  $0$  &  $0$  &  $6$  &  $1$  \\\hline\hline
\end{tabular}
\end{center}   
which lead to 
\begin{center}
\begin{tabular}{||c||c\TVR74||} \hline\hline
$({\bf a})$& $ 0\to{\widetilde V}\to {\cal O}(1,1)\oplus{\cal O}(2,1)
                                          \oplus{\cal O}(2,1)\oplus{\cal O}
                                          (3,2)$\\
           &                            $ \oplus{\cal O}(8,0)\oplus{\cal O}
                                         (4,2)\to{\cal O}(8,5)\oplus
                                         {\cal O}(12,2)\to 0$\\\hline

$({\bf b})$& $ 0\to{\widetilde V}\to {\cal O}(1,2)\oplus{\cal O}(2,0)
                                        \oplus{\cal O}(2,2)\oplus{\cal O}
                                        (3,3)$\\
           &                            $\oplus{\cal O}(8,0)\oplus
                                        {\cal O}(4,0)\to{\cal O}(8,6)
                                        \oplus{\cal O}(12,1)\to 0$\\\hline\hline
\end{tabular}
\end{center}  
The big cones of the fan $\Sigma$ corresponding to the maximal triangulation
of $\Delta$ are
\begin{eqnarray}
& &\langle e_{1}e_{2}e_{3}e_{4}\rangle,\langle e_{1}e_{2}e_{3}e_{6}\rangle,
\langle e_{1}e_{2}e_{4}e_{5}\rangle,\nonumber\\
& &\langle e_{1}e_{2}e_{5}e_{6}\rangle,\langle e_{1}e_{3}e_{4}e_{6}\rangle,
\langle e_{1}e_{4}e_{5}e_{6}\rangle,\nonumber\\  
& &\langle e_{2}e_{3}e_{4}e_{6}\rangle,\langle e_{2}e_{4}e_{5}e_{6}\rangle
\nonumber\;.  
\end{eqnarray}
The primitive collections of $\Sigma$ are $\{ e_{3},e_{5}\}$ 
and $\{e_{1},e_{2},e_{4},e_{6}\}$. Using these combinatorial data we find
$I=\langle\; z_{1}-2z_{5}-z_{6}\;,\;z_{2}-2z_{5}-z_{6}\;,\;
z_{3}-3z_{5}-z_{6}\;,\;z_{4}-4z_{5}-2z_{6}\;\rangle$ and 
$J=\langle\; z_{3}z_{5}\;,\;z_{1}z_{2}z_{4}z_{6}\;\rangle$. With
respect to the lex order $z_{1}>\ldots>z_{6}$ the Gr\"{o}bner bases
of $I+J$ and $\mbox{ann}(z_{1}+\ldots+z_{6})$ are given by
\begin{eqnarray}
I+J & = \langle & z_{1}-2z_{5}-z_{6}\;,\;z_{2}-2z_{5}-z_{6}\;,\;
                  z_{3}-3z_{5}-z_{6}\;,\;\nonumber\\
    &           & z_{4}-4z_{5}-2z_{6}\;,\; 3z_{5}^{2}+z_{5}z_{6}\;,\;
                  26z_{5}z_{6}^{3}+9
                  z_{6}^{4}\;,\;
                  z_{6}^{5}\;\rangle \nonumber\\ 
\mbox{ann}(z_{1}+\ldots+z_{6})&=\langle& z_{1}+z_{2}+z_{3}+z_{4}-11z_{5}-5z_{6}\;,\;
                                  z_{2}-2z_{5}-z_{6}\;,\;
                                  z_{3}-3z_{5}-z_{6}\;,\;\nonumber\\
                       &        & z_{4}-4z_{5}-2z_{6}\;,\;
                                  3z_{5}^{2}+z_{5}
                                  z_{6}\;,\;8z_{5}z_{6}^{2}+
                                  3z_{6}^{3}\;,\;z_{6}^{4}\;\rangle\nonumber\;. 
\end{eqnarray} 
The normalization in this case is as follows: 
\begin{eqnarray}
\langle D_{1} D_{2} D_{3} D_{6}\rangle =  
\langle D_{1} D_{2} D_{5} D_{6}\rangle = \frac{1}{2}\;\;\;,\;\;\; 
\langle D_{1} D_{2} D_{4} D_{5}\rangle = \frac{1}{3}\nonumber\;, 
\end{eqnarray}
and all other big cones
have unit volume. This results in the normalization 
$\langle z_{6}^{3}\rangle =-24$ in the intersection ring 
$A^{*}({\widetilde X})$. Therefore, we obtain
\begin{center}
\begin{tabular}{||c|c||c\TVR74||} \hline\hline
$\widetilde V$ & $c_{3}({\widetilde V})$ & $\chi({\widetilde V})$\\\hline\hline
$({\bf a})$    & $-\frac{7}{2}z_{6}^{3}$ & $\bf 42$ \\\hline
$({\bf b})$    & $z_{6}^{3}$ & $\bf -12$\\\hline\hline
\end{tabular}
\end{center}

\subsubsection*{Example 4 : ${\Bbb P}(1,1,3,3,4)$}

The gauge bundle $V$ is given by 
\begin{eqnarray}
0\to V\to {\cal O}(1)^{\oplus 2}\oplus{\cal O}(3)^{\oplus 2}
\oplus{\cal O}(8)\oplus{\cal O}(4)\to {\cal O}(8)\oplus{\cal O}(12)
\to 0\nonumber\;,
\end{eqnarray}
and the vertices 
\begin{eqnarray}
&e_{1}=(1,0,0,0), e_{2}=(0,1,0,0), e_{3}=(0,0,1,0)&\nonumber\\
&e_{4}=(0,0,0,1), e_{5}=(-1,-3,-3,-4)\;.& \nonumber
\end{eqnarray}
define the reflexive polytope $\Delta$ corresponding to 
${\Bbb P}(1,1,3,3,4)$. 
There still exists one other boundary point $e_{6}=(0,-1,-1,-1)$ of 
$\Delta$ which lies on the codimension two face spanned by $e_{1},e_{4}$
and $e_{5}$. An extra $U(1)$ 
factor arises from the desingularization. The map 
{\boldmath$\alpha$} is represented by
\begin{eqnarray}
\left(
\begin{array}{cccc}
1 & 0 & 0 & 0\\
0 & 1 & 0 & 0\\
0 & 0 & 1 & 0\\
0 & 0 & 0 & 1\\
-1&-3 &-3 &-4\\
0 & -1 & -1&-1
\end{array}
\right)\nonumber\; ,
\end{eqnarray}  
which yields 
\begin{center}
\begin{tabular}{||c||c|c|c|c|c||c\TVR74||} \hline\hline
field & $x_{1}$ & $x_{2}$ & $x_{3}$ & $x_{4}$ & $x_{5}$ & $x_{6}$\\ \hline
charge& $(1,0)$ & $(3,1)$ & $(3,1)$ & $(4,1)$ & $(1,0)$ & $(0,1)$\\\hline\hline
\end{tabular}
\end{center}     
This leads to the following equations for the data of
the resolved gauge bundle $\widetilde V$
\begin{eqnarray}
& & q_{1}+ q_{2}+ q_{3}+ q_{4}+ q_{5}+ q_{6}-p_{1}-p_{2}=0\;,\nonumber\\
& & q_{1}+ q_{2}+ 3q_{3}+ 3q_{4}+8q_{5}+ 4q_{6}-8p_{1}-12p_{2}=-26\;,\nonumber\\
& & q_{1}^{2}+ q_{2}^{2}+ q_{3}^{2}+ q_{4}^{2}+ q_{5}^{2}+ q_{6}^{2}
-p_{1}^{2}-p_{2}^{2}=-6\;.\nonumber
\end{eqnarray} 
Two solutions of these equations are
\begin{center}
\begin{tabular}{||c|c|c|c|c|c||c|c||} \hline\hline
$q_{1}$ &$q_{2}$ &$q_{3}$ &$q_{4}$ &$q_{5}$ &$q_{6}$ &$p_{1}$ &$p_{2}$\\\hline
  $1$  &  $1$  &  $0$  &  $0$  &  $2$  &  $1$  &  $3$  &  $2$  \\\hline
  $2$  &  $0$  &  $0$  &  $0$  &  $3$  &  $1$  &  $4$  &  $2$  \\\hline\hline
\end{tabular}
\end{center}   
which result in 
\begin{center}
\begin{tabular}{||c||c\TVR74||} \hline\hline
$({\bf a})$& $ 0\to{\widetilde V}\to {\cal O}(1,1)\oplus{\cal O}(1,1)
                                          \oplus{\cal O}(3,0)\oplus{\cal O}
                                          (3,0)$\\
           &                            $ \oplus{\cal O}(8,2)\oplus{\cal O}
                                         (4,1)\to{\cal O}(8,3)\oplus
                                         {\cal O}(12,2)\to 0$\\\hline

$({\bf b})$& $ 0\to{\widetilde V}\to {\cal O}(1,2)\oplus{\cal O}(1,0)
                                         \oplus{\cal O}(3,0)\oplus{\cal O}
                                        (3,0)$\\
           &                           $\oplus{\cal O}(8,3)\oplus
                                       {\cal O}(4,1)\to{\cal O}(8,4)
                                       \oplus{\cal O}(12,2)\to 0$\\\hline\hline
\end{tabular}
\end{center}  
The big cones of the fan $\Sigma$ corresponding to the maximal triangulation
of $\Delta$ are
\begin{eqnarray}
\langle e_{1}e_{2}e_{3}e_{4}\rangle,\langle e_{1}e_{2}e_{3}e_{5}\rangle,
\langle e_{1}e_{2}e_{4}e_{6}\rangle,\nonumber\\
\langle e_{1}e_{2}e_{5}e_{6}\rangle,\langle e_{1}e_{3}e_{4}e_{6}\rangle,
\langle e_{1}e_{3}e_{5}e_{6}\rangle,\nonumber\\  
\langle e_{2}e_{3}e_{4}e_{5}\rangle,\langle e_{2}e_{4}e_{5}e_{6}\rangle,
\langle e_{3}e_{4}e_{5}e_{6}\rangle\nonumber\;.  
\end{eqnarray}
The primitive collections of $\Sigma$ are $\{ e_{2},e_{3},e_{6}\}$ 
and $\{e_{1},e_{4},e_{5}\}$. From these combinatorial data we find
$I=\langle\; z_{1}-z_{5}\;,\;z_{2}-3z_{5}-z_{6}\;,\;
z_{3}-3z_{5}-z_{6}\;,\;z_{4}-4z_{5}-z_{6}\;\rangle$ and 
$J=\langle\; z_{1}z_{4}z_{5}\;,\;z_{2}z_{3}z_{6}\;\rangle$. With
respect to the lex order $z_{1}>\ldots>z_{6}$ the Gr\"{o}bner bases
of $I+J$ and $\mbox{ann}(z_{1}+\ldots+z_{6})$ are given by
\begin{eqnarray}
I+J & = \langle & z_{1}-z_{5}\;,\;z_{2}-3z_{5}-z_{6}\;,\;
                  z_{3}-3z_{5}-z_{6}\;,\;z_{4}-4z_{5}-z_{6}\;,\;\nonumber\\
    &           & 4z_{5}^{3}+z_{5}^{2}z_{6}\;,\;9z_{5}^{2}z_{6}+6z_{5}
                  z_{6}^{2}+z_{6}^{3}\;,\;18z_{5}z_{6}^{3}+5z_{6}^{4}\;,\;
                  z_{6}^{5}\;\rangle \nonumber\\ 
\mbox{ann}(z_{1}+\ldots+z_{6})&=\langle& z_{1}+z_{2}+z_{3}+z_{4}-11z_{5}-3z_{6}\;,\;
                                  z_{2}-3z_{5}-z_{6}\;,\;z_{3}\nonumber\\
                       &        & -3z_{5}-z_{6}\;,\;
                                  z_{4}-4z_{5}-z_{6}\;,\;4z_{5}^{3}+z_{5}^{2}
                                  z_{6}\;,\;3z_{5}z_{6}+
                                  z_{6}^{2}\;,\;z_{6}^{4}\;\rangle\nonumber\;. 
\end{eqnarray} 
The normalization in this case is as follows: $\langle D_{1} D_{2}
D_{3} D_{5}\rangle = \frac{1}{4}$ and all other big cones
have unit volume. This leads to the normalization 
$\langle z_{6}^{3}\rangle =36$ in the intersection ring 
$A^{*}({\widetilde X})$. Therefore, we obtain
\begin{center}
\begin{tabular}{||c|c||c\TVR74||} \hline\hline
$\widetilde V$ & $c_{3}({\widetilde V})$ & $\chi({\widetilde V})$\\\hline\hline
$({\bf a})$    & $-\frac{26}{9}z_{6}^{3}$ & $\bf -52$ \\\hline
$({\bf b})$    & $-\frac{20}{9}z_{6}^{3}$ & $\bf -40$\\\hline\hline
\end{tabular}
\end{center}


\section{Conclusion}

Starting from a series of solutions of the anomaly cancellation 
equation we have constructed a class of $(0,2)$ Calabi-Yau $\sigma$ 
models. These solutions are associated
to certain $(0,2)$ Landau-Ginzburg models which are conjectured to be 
`equivalent' to the $(0,2)$ superconformal field theories constructed in
\cite{bl961,nik961,nik962}. Following \cite{di96} we have studied the 
desingularization of a few examples from this class. This led in each
case to a family of $(0,2)$ Calabi-Yau $\sigma$ models. 
As pointed out above, the ambiguity in the geometric
interpretation of the $(0,2)$ models has, in contrast to the $(2,2)$ case,
two different sources. The first one is, as in $(2,2)$ models, the choice of
maximal triangulations of the reflexive polytope $\Delta$ \cite{as93,as941}. 
The second one comes from the different ways of `pulling the 
gauge bundle back to the
desingularized Calabi-Yau variety'. 
It seems to be natural to ask if
there exists a selection rule which associate to a given superconformal
field theory a subset of the desingularized Calabi-Yau $\sigma$ models
as its possible geometric realizations. The explicit knowledge of the 
exact superconformal theories in our case will be useful in answering
this question for the class of models considered here. 
Another issue that can be addressed is the following.
As mentioned above, the solution of the anomaly
cancellation equation can be equally interpreted as the defining data
of a bundle on a Calabi-Yau complete intersection. (To deal with this
latter case is, however, technically more cumbersome.) 
It would be interesting to study  the consequences
of this dual interpretation of the anomaly cancellation condition
and to compare the desingularized models in these two cases.\\

{\bf Acknowledgements}: I would like to thank M. Kreuzer and R. Blumenhagen
for helpful discussions and comments. This work has been supported by 
the {\em Austrian Research Fund} (FWF) under grant Nr. P10641-PHY.


\newpage

\subsection*{Appendix : Gr\"{o}bner basis}

Let $k[x_{1},\ldots,x_{m}]$ denote a polynomial ring in $m$ variables
$x_{1},\ldots,x_{m}$ over the field $k$. To each monomial $x^{\alpha}=
x_{1}^{\alpha_{1}}\ldots x_{m}^{\alpha_{m}}$ we associate the element
$(\alpha_{1},\ldots,\alpha_{m})$ in the semigroup $({\Bbb N}^{m}, +)$. By
a {\em monomial ordering} in $k[x_{1},\ldots,x_{m}]$ we mean an order relation 
on the set of monomials induced by a total ordering $>$ on ${\Bbb N}^{m}$
which is consistent with its semigroup structure and such that $>$ is 
a well-ordering on ${\Bbb N}^{m}$. Here are two examples: (1) {\em 
lex\/}(icographic) order: $x^{\alpha}>_{lex}x^{\beta}\; 
:\Leftrightarrow$ (the left-most nonzero entry in $\alpha -\beta$ is positive),
(2) {\em g\/}(raded) {\em lex}  order: $x^{\alpha}>_{glex}x^{\beta}\; 
:\Leftrightarrow (\sum_{i}\alpha_{i}>\sum_{i}\beta_{i}) $ or $ (\alpha =
\beta$ and $x^{\alpha}>_{lex}x^{\beta})$. Given a nonzero polynomial
$f=\sum_{\alpha}a_{\alpha}x^{\alpha}$ in $k[x_{1},\ldots,x_{m}]$ with a
monomial order we define: $\mbox{deg}(f)=\alpha_{\tt max}= \mbox{max}
(\alpha\in{\Bbb N}^{m}: a_{\alpha}\not= 0)$, leading term of $f= lt(f)=
a_{\alpha_{\tt max}}x^{\alpha_{\tt max}}$, leading monomial of $f= lm(f)=
x^{\alpha_{\tt max}}$. We now come to the division algorithm. 

\vspace{2mm}
\noindent
{\itshape\bfseries Division algorithm in} $k[x_{1},\ldots,x_{m}]$

\vspace{1mm}
\noindent
\hspace*{5mm}{\em input\/}: a $s$-tuple of polynomials 
             $F=(f_{1},\ldots,f_{s})$ and a nonzero polynomial $f$ ,\\
\hspace*{5mm}{\em output\/}: the reminder $r\; (=\bar{f}^{\tt F})$ of 
            dividing $f$ by $F$ and the quotients $q_{1},\ldots,q_{s}$,\\
\hspace*{5mm}{\em algorithm\/}: $p:=f\;,\;r:=0\;,\;q_{i}:=0 $ 
            for all $i=1,\ldots,s$\\
\hspace*{22mm}{\itshape\bfseries repeat} \\
\hspace*{30mm}$i:=1$ , dividing:=true\\
\hspace*{30mm}{\itshape\bfseries while} $ (i\leq s) $ and (dividing) 
              {\itshape\bfseries do}\\
\hspace*{30mm}{\itshape\bfseries if} $\; lt(f_{i})\; $ divides 
               $ \;lt(p)\; $ {\itshape\bfseries then}\\ 
\hspace*{35mm}$u:=lt(p)/lt(f_{i})\;,\;q_{i}:=q_{i}+u\;,\;p:=p-u\,f_{i}\;,\;$
              dividing:=false\\
\hspace*{30mm}{\itshape\bfseries else} $ \;i:=i+1 $\\
\hspace*{30mm}{\itshape\bfseries if} dividing {\itshape\bfseries then}\\
\hspace*{35mm}$r:=r+lt(p)\;,\;p:=p-lt(p)$\\
\hspace*{22mm}{\itshape\bfseries until} $ p=0$     

\vspace{1mm}
\noindent
It should be noted that, for a given monomial order, $r$ and $q_{i}$ depened on
the order of $f_{i}$ in $F$. Given $f,g\in k[x_{1},\ldots,x_{m}]$ with
$h=\mbox{LCM}(lm(f),lm(g))$ we define the $S$-polynomial of $f$ and $g$ as
$S(f,g)=h\cdot (f/lt(f)-g/lt(g))$. Now let $I$ be an ideal in 
$k[x_{1},\ldots,x_{m}]$. A {\em Gr\"{o}bner basis} of $I$ is a generating
set $G=\{f_{1},\ldots,f_{s}\}$ such that $\overline{S(f_{i},f_{j})}
^{\;\tt G}=0$ for all $i$ and $j$. The reminder of dividing a polynomial by 
$G$ is unique! Using the {\em Buchberger's algorithm} one can find a 
Gr\"{o}bner basis of a given ideal.
 
\vspace{2mm}
\noindent
{\itshape\bfseries Buchberger's algorithm } 

\vspace{1mm}
\noindent
\hspace*{5mm}{\em input\/}: a $s$-tuple of polynomials 
             $F=(f_{1},\ldots,f_{r})$ which generates $I$ ,\\
\hspace*{5mm}{\em output\/}: a Gr\"{o}bner basis $G=(g_{1},\ldots,g_{s})$ 
             of $I$,\\
\hspace*{5mm}{\em algorithm\/}: $G:=F$ \\
\hspace*{22mm}{\itshape\bfseries repeat} \\
\hspace*{30mm}$G':=G$\\
\hspace*{30mm}{\itshape\bfseries for} each $i\;,\;j$ with 
               $i\not = j \; $ in $G' $ {\itshape\bfseries do} \\
\hspace*{30mm}$S:=\overline{S(f_{i},f_{j})}^{\;\tt G'}$\\
\hspace*{30mm}{\itshape\bfseries if} $ S\not = 0 $ 
              {\itshape\bfseries then} $ G:=G\cup \{S\}$ \\ 
\hspace*{22mm}{\itshape\bfseries until} $ G=G'$     

\vspace{1mm}
\noindent
Using the Gr\"{o}bner basis we can do algorithmic calculations in a
polynomial ring. As an example we give the algorithm for the calculation of
$I:J=\{f\in k[x_{1},\ldots,x_{m}] \mid fJ\subset I\}$. First we determine a
Gr\"{o}bner basis of $I\cap J$. It is given as the intersection of
$k[x_{1},\ldots,x_{m}]$ with a 
Gr\"{o}bner basis of the ideal $tI-(1-t)J$ in $k[t,x_{1},\ldots,x_{m}]$
with respect to a lex order in which $t$ is greater than $x_{i}$. 
Let $J=\langle f_{1},\ldots,f_{n}\rangle$. Taking
$I:J=I:\langle f_{1},\ldots,f_{n}\rangle=\bigcap_{i=1}^{n}I:\langle 
f_{i}\rangle $ into account we only need to calculate a  
Gr\"{o}bner basis of $I:\langle f_{i}\rangle $. Because of
$I:\langle f_{i}\rangle =1/f(I\cap\langle f_{i}\rangle$ it reduces to the 
case just discussed above (cf. \cite{ei95,ad94,cox921} for more details). 
    




\begin{thebibliography}{11}

\addtolength{\itemsep}{-20pt} \small \vspace{2mm}

\bibitem{ba88} T.Banks, L.J.Dixon, D.Friedan, E.Martinec, {\em Phenomenology 
               and conformal field theory  or can string theory 
               predict the weak mixing angle?,} \npb 299 (1988) 613 \\
\bibitem{ge88}  D.Gepner, {\em Space-time supersymmetry in compactified 
                string theory and superconformal models,} \npb 296 (1988) 757;
                {\em $N=2$ string theory}, in: Proceedings of the Trieste 
                Spring School Strings 1989, eds. M.Green et al. , 
                World Scientific, Singapore, 1990       \\
\bibitem{hu85} C.M.Hull, E.Witten, {\em Supersymmetric sigma models 
               and the heterotic string,} \plb 160 (1985) 398   \\
\bibitem{wi86} E.Witten, {\em New issues in manifolds of SU(3) holonomy,} 
               \npb 268 (1986) 79 \\
\bibitem{di86}  M.Dine, N.Seiberg, X.G.Wen, E.Witten, {\em Nonperturbative 
                effects on the string worldsheet I and II,} 
                \npb 278 (1986) 769, {\bf B289} (1987) 319 \\
\bibitem{si95}  E.Silverstein, E.Witten, {\em Criteria for conformal 
                invariance of (0,2) models,} \npb 444 (1995) 161, 
                hep-th/9503212 \\
\bibitem{di871} J.Distler, {\em Resurrecting (2,0) compactifications,}
                \plb 188 (1987) 431     \\  
\bibitem{di881} J.Distler, B.Greene, {\em Aspects of (2,0) string 
                compactifications,} \npb 304 (1988) 1   \\
\bibitem{gr90}  B.R.Greene, {\em Superconformal compactifications in 
                weighted projective space,} \cmp 130 (1990) 335 \\
\bibitem{wi93}  E.Witten, {\em Phases of N=2 theories in two dimensions,} 
                \npb 403 (1993) 159, hep-th/9301042 \\
\bibitem{di94}  J.Distler, S.Kachru, {\em (0,2) Landau--Ginzburg theory,}
                \npb 413 (1994) 213, hep-th/9309110     \\
\bibitem{di95}   J. Distler, {\em Notes on (0,2) superconformal field
                 theories,} Proceedings of the 1994 Trieste Summer School,
                 hep-th/9502012 \\
\bibitem{bl95}  R.Blumenhagen, A.Wi{\ss}kirchen, {\em Exactly solvable (0,2) 
                supersymmetric string vacua with GUT gauge groups,} 
                \npb454 (1995) 561, hep-th/9506104;\\
                {\em Exploring the moduli space of (0,2) strings,} 
                \npb 475 (1996) 225, hep-th/9604140 \\
\bibitem{bl96}  R.Blumenhagen, R.Schimmrigk, A.Wi{\ss}kirchen, {\em The (0,2) 
                exactly solvable structure of chiral rings, Landau--Ginzburg 
                theories and Calabi--Yau manifolds,} \npb 461 (1996) 
                460, hep-th/9510055     \\
\bibitem{bl961} R.Blumenhagen, R.Schimmrigk, A.Wi{\ss}kirchen, 
                {\em (0,2) mirror symmetry,} hep-th/9609167     \\
\bibitem{nik961} M. Kreuzer, M. Nikbakht--Tehrani, {\em (0,2) string
                compactifications,} hep-th/9611130 \\
\bibitem{nik962} M. Kreuzer, M. Nikbakht--Tehrani, in preparation \\
\bibitem{di96}  J.Distler, B.R.Green, D.R.Morrison, {\em Resolving 
                singularities in (0,2) models,} hep-th/9605222  \\
\bibitem{di951}  J. Distler, S. Kachru, {\em Duality of (0,2) string vacua,}
                 \npb 442 (1995) 64, hep-th/9501111 \\
\bibitem{DA78}  V.I.Danilov, {\em The geometry of toric varieties,} 
                Russian Math. Survey {\bf33}, n.2 (1978) 97     \\
\bibitem{FU93}  W.Fulton, {\em Introduction to toric varieties,} 
                Princeton Univ. Press, Princeton, 1993  \\
\bibitem{OD88}  T.Oda, {\em Convex bodies and algebraic geometry,} 
                Springer Verlag, 1988   \\
\bibitem{cox92}  D. Cox, {\em The homogeneous coordinate ring of a toric
                 variety,} alg-geom/9210008 \\
\bibitem{fl89}  A.R.Fletcher, {\em Working with complete intersections,} 
                Bonn preprint MPI/89--35 (1989) \\ 
\bibitem{bo93}  L.A.Borisov, {\em Towards the mirror symmetry for Calabi-Yau
                complete intersections in Gorenstein toric fano varieties,}
                alg-geom/9310001        \\
\bibitem{ba95}  V.V.Batyrev, L.A.Borisov, {\em Dual cones and mirror 
                symmetry for generalized Calabi--Yau manifolds,} 
                alg-geom/9402002;
                {\em On Calabi-Yau complete intersections in toric varieties,}
                alg-geom/9412017 \\
\bibitem{ba94}  V.V.Batyrev, {\em Dual polyhedra and mirror symmetry 
                for Calabi--Yau hypersurfaces in toric varieties,} 
                J. Alg. Geom. {\bf 3} (1994) 493, alg-geom/9310003      \\
\bibitem{mo95}  D.R.Morrison, M.R.Plesser, {\em Summing the instantons: 
                quantum cohomology and mirror symmetry in toric 
                varieties,} \npb 440 (1995) 279, hep-th/9412236 \\
\bibitem{as941}  P.S. Aspinwall, B.R. Greene, D.R. Morrison, {\em Calabi-Yau
                 moduli space, mirror manifolds and spacetime topology
                 change in string theory,} \npb 416 (1994) 414, 
                 hep-th/9309097\\
\bibitem{as94}  P.S.Aspinwall, B.R.Greene, D.R.Morrison, {\em Space-time 
                topology change and stringy geometry,} 
                \jmp 35 (1994) 5321     \\
\bibitem{as942}  P.S. Aspinwall, B.R. Greene, {\em On the geometric 
                 interpretation  of $N=2$ superconformal theories,}
                 \npb 437 (1995) 205, hep-th/9409110 \\
\bibitem{as93}  P.S.Aspinwall, B.R.Greene, D.R.Morrison, {\em Multiple mirror
                manifolds and topology change in string theory,} 
                \plb 303 (1993) 249, hep-th/9301043     \\
\bibitem{ba93}   V.V. Batyrev, {\em Quantum cohomology rings of toric 
                 manifolds,} Journ\'{e}es de G\'{e}om\'{e}trie Alg\'{e}brique
                 d'Orsay (Juillet 1992), Ast\'{e}risque, Vol. 218, 
                 Soci\'{e}t\'{e} Math\'{e}matique de France, 1993, p. 9,
                 alg-geom/9310004 \\
\bibitem{ei95}   D. Eisenbud, {\em Commutative algebra with a view 
                 toward algebraic geometry,} GTM 150, Springer
                 Verlag, 1995 \\
\bibitem{ad94}   W. W. Adams, P. Loustaunau, {\em  An introduction to 
                 Gr\"{o}bner bases,} Graduate Studies in Math., Vol. 3,
                 AMS, 1994 \\
\bibitem{cox921} D. Cox, J. Little, D. O'Shea, {\em  Ideals, varieties, 
                 and algorithms: an introduction to computational algebraic 
                 geometry and commutative algebra,} UTM, 
                 Springer Verlag, 1992  \\ 
\end{thebibliography}
\end{document}